%% file: tor.tex
\documentclass[%
final,
reprint,
notitlepage,
narroweqnarray,
inline,
twoside,
]{ieee}

\usepackage{url}
\usepackage{amsmath}
\usepackage{pifont}
\newcommand{\cmark}{\ding{51}}%
\newcommand{\xmark}{\ding{55}}%
\graphicspath{{img/}}
\usepackage{algorithm}
\usepackage{algpseudocode}
\usepackage{ulem}

\begin{document}

\title{A Composite-Metric Based Path Selection Technique for the Tor Anonymity Network}

\author[Momeni AND Kharrazi]{Sadegh Momeni Milajerdi,\authorinfo{S.\,Momeni Milajerdi is with the Department of Computer
Engineering, Sharif University of Technology, Tehran, Iran.
Phone: $+$\,98\,912\,596--7299, e-mail: smomeni@ce.sharif.edu}%
\and{}and Mehdi Kharrazi\authorinfo{M.\
Kharrazi is with the Department of Computer
Engineering, Sharif University of Technology, Tehran, Iran.
Phone: $+$\,98\,21\,6616--6627, 
e-mail: kharrazi@sharif.edu}
}




\maketitle 

\begin{abstract} 


The Tor anonymous network has become quite popular with regular users on the Internet. In the Tor network, an anonymous path is created by selecting three relays through which the connection is redirected. Nevertheless, as the number of Tor users has increased substantially in recent years, the algorithm with which the relays are selected affects the performance provided by the Tor network. More importantly as the performance suffers, users will leave the network, resulting in a lower anonymity set and in turn lower security provided by Tor network. 

In this paper, we proposed an algorithm for improving performance and security of the Tor network, by employing a combination of different metrics in the process of the path selection between the source and destination node. These metrics are bandwidth and uptime of relays as node conditions and delays between the relays as a path condition. Through a number of experiments we show that we could double the performance observed by end users when using the proposed technique as opposed to the current Tor path selection algorithm. More importantly, the proposed technique only requires a software upgrade on the client side, and other Tor nodes do not need to be modified. 

%
%
\end{abstract}


\section{Introduction}

\PARstart Tor~\cite{tor-design} is the most well-known anonymous network which has a large user-base around the globe. As a general rule, when the number of participants increases in an anonymous network, the overall anonymity provided by the network also increases. The rapid growth of the Tor network has not been anticipated in its design, which has led to many latency issues. For example in a world scale network, the Tor protocol may at times circulate packets several times around the world before delivering them to the receiver; or the increase in the number of nodes connecting and directing traffic through the Tor, has resulted in Tor relays being overloaded. This effect has not been small. McCoy et al.~\cite{mccoy-pet2008} note  a five-fold increase in latency experienced by clients as compared to when their connection does not go through the Tor network. This increase in latency leads to user dissatisfaction and a reduction in the number of active users; resulting in a decrease in the anonymity provided by the network.


Anonymity is provided in the Tor network, by redirecting the connection through multiple Tor relays (usually three relay nodes are used), before the connection is made to the final destination. A list of available nodes, willing to operate as relays, with their bandwidth and uptime statistics is published by Tor directory servers. Based on the published information, a Tor client selects the nodes through which the Tor path is created.

The proper selection of these relays, directly affects the performance observed by the communicating parties. Where in general, the overall delay experienced by the client would consist of transmission delay (i.e. link bandwidth), queuing delay (i.e. due to network congestion), and propagation delay (i.e. physical distance between nodes). For example if low bandwidth relays are selected, the client will experience a low bandwidth connection, and is more likely to leave the network.


Nevertheless, always selecting relays with higher bandwidth and larger uptime values creates other problems. In fact, there is an important trade off when selecting the Tor relays.  On one side, and from the performance aspect, a high bandwidth node with a large uptime value (i.e. reliability of the node) would be preferential. But, from a security perspective, if nodes are chosen in a deterministic manner (i.e. always best available node) then an attacker could, with high probability, guess which nodes will be selected by a client and in turn mount an attack on the anonymous network. Hence, there should be some randomness in selecting the nodes.

Other than bandwidth and uptime, a third parameter which effects the performance obtained by a client is path latency. There have been a number of proposals in~\cite{DBLP:conf/pet/SherrBL09,panchenko2009path,congestion-tor12}, which measure the path delay, and consider it as a factor when selecting the Tor relays in the process of creating a connection through the network. An alternate approach is proposed by Akhoondi et al~\cite{oakland2012-lastor}, where they argue that the geographical location of nodes could be used as a good measurement base to calculate the path latency between them; where this information could be obtained by using available IP to geographical location mappings. 

In this work, we proposed a new path selection methodology which takes into account the delay between relays, in addition to the bandwidth and uptime of the relays , when selecting them to create an anonymous path. Our approach differs from previous works as it exploits composite-metrics for path selection, considers usage history of users and so incurs low overhead on the resources and minimum run-time to provide a suitable path. More specifically our contributions are:

\begin{itemize}

\item Proposing a path selection algorithm which efficiently selects a set of relays for an anonymous connection while considering bandwidth, reliability, and delay between the selected relays for improved end-to-end performance.
\item Evaluating the proposed technique through a number of experiments, in which it is shown that the proposed technique could double the performance observed by the client as compared to the default path selection technique used in the Tor network. 
\item By creating paths to the average geo-location point visited by the client, the proposed algorithm avoids creating a new path for each new destination and hence  more stable paths are created and the client does not get delayed for each new destination he/she intends to visit.
\item The proposed algorithm requires no change to the current Tor network and changes are only required on the clients which are interested in obtaining improved performance. 
\end{itemize}

In the rest of this manuscript, we discuss related works in Section~\ref{rel}. In Section ~\ref{geo}, we introduce our proposed algorithm for path selection in the Tor network and in Section~\ref{eval} we evaluate the proposed method and compare the results with different path selection algorithms. We discuss the obtained results and how it compares with related work in this area in Section~\ref{dis}. Finally, we conclude in Section~\ref{con}.


\section{related works}\label{rel}

There have been a number of previous studies on improving path selection in the Tor anonymous network.  A number of studies~\cite{feamster:wpes2004,DBLP:conf/ccs/EdmanS09} select the best path with respect to the Autonomous Systems traversed. 
A path selection method in which overloaded nodes are avoided is proposed \cite{ipccc12-performance}. Sherr et al~\cite{sherr2014design} introduce a flexible path selection design in which applications select a trade-off between performance and anonymity based on the user's specific requirements. Furthermore the notion of trust is employed in ~\cite{ccs2011-trust,johnson2009more,sassone2010trust,asiaccs2014-reputation} in order to increase security. In addition~\cite{ccs2013-pctcp} propose a method for improving performance of Tor by changing its transport layer design. There are also a number of proposals~\cite{DBLP:conf/pet/SherrBL09,panchenko2009path,congestion-tor12,oakland2012-lastor} with the goal of optimizing the path while considering the latency between relays. 

Wacek et al.~\cite{ndss13-relay-selection} compare and contrast a number of path selection algorithms through a number of experiments. They observe that ignoring the relay bandwidths when selecting a path, results in an overall poor performance. Furthermore, considering distance between relays would improve the end-to-end delays. In the rest of this section we will focus on techniques which consider the latency between relays and consider the rest of the approaches as out of scope.



Sherr et al. \cite{DBLP:conf/pet/SherrBL09} propose a method in which every node in the network is disclosed as a point in an n-dimensional coordinate system. RTT between two nodes is attributed to distance of related two points in the coordinate system. Each node calculates its exact place in the system using the ping protocol and through a united coordinate system downloaded through the directory server could estimate the RTT between available nodes. Nevertheless, the proposed method requires a software update to all Tor clients and relays, which is not very practical. Moreover, there is no mechanism considered for prohibiting relays from announcing wrong information about their place which could lead to an incorrect coordinate system. 


In order to mitigate the above noted problems, Panchenko et al.~\cite{panchenko2009path} propose that each client which wants to calculate RTT between relays, should first create a path using those relays and then force the last relay to send a packet to the sender with which the RTT could be measured. Where this could be achieved by transmitting a packet which violates the exit policy of last relay, in turn compelling it to transmit a feedback packet to the initial sender. Nevertheless, such approach would create a lot of excessive traffic and is time consuming for measuring RTTs.

An alternate approach is proposed in~\cite{congestion-tor12}, in which generation of excessive traffic is avoided. More specifically, in Tor special control packets are generated intermittently (i.e. when a new TCP exit stream is established, after a certain number of data cells are transmitted, or etc.). Hence by observing these packets, the sender could estimate the RTT of the path. Although this method has no extra load on the network traffic, it requires a long execution time before it could provide a good view of the latencies in the network.

Akhoondi et al.~\cite{oakland2012-lastor}  propose a practical, low overhead, and reliable method for calculating RTT between Tor relays. They argue that geographical distance between nodes is a good indicator for latency between the nodes. Therefore, they argue that the distances between geographical location of nodes on a 2-dimensional map is a good estimate for RTT between them. In order to avoid high computation cost, by considering all possible nodes for calculating path latencies, the nodes geographically close to each other are grouped into a set of clusters, and then a random node is chosen from each selected cluster, where clusters are selected so that the end-to-end geographical distance is minimized.  Although such approach decreases the run-time of the algorithm, but it also reduces the precision in selecting shorter paths. Furthermore, by clustering many  geographically close nodes, one is unable to select the better node in the cluster based on other factors such as the node's bandwidth or uptime.

Another problem is that their method is dependent to the location of specific destination in which a user wants to make a connection with. So path selection should be done after the destination server has been determined, which leads to putting users on hold while the circuit to that destination is established. 


Altogether by only reducing propagation delay, one cannot make a comprehensive path selection method. Because a large share of delay is related to bandwidth and reliability of relays participating in a circuit. So a comprehensive path selection algorithm should consider different aspects of network delay (i.e. propagation delay, queuing delay, and transmission delay). In what follows we will propose a new approach in which all these parameters are considered.

\section{proposed algorithm}\label{geo}

As only a small percentage of Tor relays have very high bandwidths, path selection approaches which only consider relay bandwidths would tend to select such nodes, creating more deterministic paths and also congesting these few nodes by placing heavy loads on them. On the other hand, only selecting relays based on their geographical location,  will put heavier load on some relays near specific high traffic locations.  Hence, it would be beneficial to select paths while considering three parameters, up-time for reliability, bandwidth for performance, and geographical locations for latency. 

An important observation is that users generally visit some specific websites on a daily basis, checking email, using some particular social network, or visiting news websites. This deterministic pattern enables one to select the Tor path which ends closest to the areas regularly visited by the client. Hence, by storing geographical location data of the servers visited by the user and the frequency of the visits, one could obtain an average geo-location point by averaging the vertical and horizontal positions of the visited servers.

For instance if someone accesses a web server in China two times and a web server in England three times, then the average-geo-location point will be on the length of $\dfrac{2}{5}$ on the link connecting China to England. Based on this simple heuristic our path selection method will be independent of the destination address. This independence brings important benefits to the proposed approach such as more stable and permanent circuits, lower traffic overhead, and lower setup time. These benefits will be discussed in detail in Section~\ref {dis}.

In the proposed path selection technique relays in a path are chosen incrementally and one after the other. For example, in a circuit consisting of three relays, first we choose the last relay, then the second one, and then the first relay in the path. In each phase we consider all the relays available in the Tor directory server and each one gets a rank based on its bandwidth, uptime, and location. This method is independent to the destination address which user intends to make a connection with, instead we consider an average-geo-location point as noted above.

We use different functions for calculating latency of nodes in each step of our method. The third relay's latency is measured given the average-geo-location of the servers visited by the client. Afterwards, the second relay's latency is measured based on the third relay's location, and the first relay's latency in the path is measured based on the second relay's location. In what follows we first present the methodology in which the distance  between the client and avg. geo-location is measured and then later in the section we discuss how relays are ranked and selected based on the discussed parameters (i.e. delay, bandwidth, and reliability).





\begin{algorithm}
\caption{Latency Measuring pseudo-code}
\label{alg1}
\begin{algorithmic}
\Procedure {selectPath}{$S$,$H$,$D$}
\State \textbf{inputs} $= source~s, History~H, Directory~D$
\State \textbf{list} $circuit$
\State \textbf{node} $R1,R2,R3$
\State $(x,y,sum,Avg) \gets 0$
\Comment{selecting R3:}
\If {$H~is~Empty$}
\State $x \gets Random[0,1)$
\State $R3 \gets D.element(x*D.size)$
\Else
\ForAll {$nodes~n \in H$}
\State $Avg_x \gets Avg_x+n.x*n.count$
\State $Avg _y \gets Avg_y+n.y*n.count$
\State $sum \gets sum+n.count$
\EndFor
\State $Avg _x \gets Avg_x/sum$
\State $Avg_y \gets Avg_y/sum$
\ForAll {$nodes~n \in D$}
\State $n.RD \gets dist(s,n) + dist(Avg,n)$
\EndFor
\EndIf
\State $select~R3~after~computing~rank~of~all~nodes$
\ForAll {$nodes~n \in D$}
\Comment{selecting R2:}
\State $n.RD \gets dist(s,n) + dist(R3,n)$
\EndFor
\State $select~R2~after~computing~rank~of~all~nodes$
\ForAll {$nodes~n \in D$}
\Comment{selecting R1:}
\State $n.RD \gets dist(s,n) + dist(R2,n)$
\EndFor
\State $select~R1~after~computing~rank~of~all~nodes$
\EndProcedure
\end{algorithmic}
\end{algorithm}

\subsection{Relative Distance}\label{dist}

We defined a Relative Distance measure, in short $RD$, with the goal of minimizing the end-to-end delay in the selected paths. As noted in  \cite{oakland2012-lastor}, 2D geographical distance of two nodes on a map can be used as an estimate for the RTT between them.  Although packets do not always traverse in direct paths and the traversed path depends on the underlying network topology, nevertheless minimizing the distance between relays results in a decreased delay. Hence we calculate the distance between two nodes, $a$ and $b$, in a 2D space by considering their x-y coordinates and using the noted relationship:
\begin{eqnarray*}
\label{eqn:dist}
dist(a,b) = \sqrt{(x_a-x_b)^2+(y_a-y_b)^2}
\end{eqnarray*}

Afterwards, we define $RD$ as a measure in relation to the source node's geographical location. More specifically, we defined the $RD$ of node A as equal to the distance of the source node to the destination node when the path goes through node A. For example, the $RD$ value for the third relay (i.e. R3, the last relay used in the path) is calculated by:

\begin{eqnarray*}
RD_{R3}(n) = dist(n,source\ node)+dist(n,Avg\_Geo)
\end{eqnarray*}

We then calculate the $RD$ for the second relays, R2,  as from the source to the third relay, R3: 
\begin{eqnarray*}
RD_{R2}(n) = dist(n,source\ node)+dist(n,relay_3)
\end{eqnarray*}

Similarly, the $RD$ for the first relay, R1, is calculated by:
\begin{eqnarray*}
RD_{R1}(n) = dist(n,source\ node)+dist(n,relay_2)
\end{eqnarray*}
The pseudo-code of the algorithm for measuring latency for each relay node is available in \ref{alg1}.

\subsection{Relay Ranking and Selection}\label{rank}

As noted earlier, the proposed technique considers three metrics of geographical location, bandwidth, and reliability, when selecting the relays. Therefore, we need to assign a ranking score to each relay in order to select the proper relay at each step. After comparing many different combinations we believe the following relationship gives proper weight to each of the three considered parameters:
\begin{eqnarray*}
\label{eqn:rank}
Rank_{n}=perf_{n}*(1-latency_{n})*\log_2{(1+reliability_{n})}
\end{eqnarray*}
Where, $Rank_{n}$ designates the rank assigned to node $n$, in which:
\begin{eqnarray*}
perf_{n} & = & \frac{bandwidth_{n}}{max\ bandwidth} \\
latency_n & = & \frac{RD_n}{max\ RD} \\
reliability_n & = & \frac{uptime_n}{max\ uptime}
\end{eqnarray*}
For each node $performance_{n}$ is its bandwidth divided by maximum bandwidth in the network and $latency_{n}$ as the RD of that node to the maximum available RD. Also we construct $Reliability_{n}$ as the uptime of the node in seconds divided by maximum available uptime in the network. Furthermore and in order to make a clear distinction between low uptime values and higher uptime values, we weight the uptime value by a logarithm function in the ranking equation noted above. This could be explained by the fact that high uptime relays are almost as reliable as each other but low uptime relays should be differentiated strictly.


After ranking the relays each relay with higher rank value has a greater chance for being selected. Therefore each node which has lower latency (distance to a preferred location), higher bandwidth, and higher uptime has a greater rank. But, and as noted earlier, it is not possible to always select the highest ranking relay as the created path should be somewhat unpredictable, otherwise an attack could be mounted against the anonymity network.

To that end, we employ an approach noted in~\cite{panchenko2009path} with which some randomness is introduced when selecting relays in order to prevent an attacker from anticipating the path that will be used.
More specifically, a random number is selected from zero to sum of all the ranks. Then starting from the highest ranked relay, relay rankings are added together until this amount gets bigger than the random value, at which time the last counted relay will be selected. Hence each relay has some probability according to its rank for being selected. The pseudo-code of the algorithm is presented as Algorithm~\ref{alg2}.

\begin{algorithm}
\caption{relay selection pseudo-code}
\label{alg2}
\begin{algorithmic}
\Procedure {selectbyRank}{$D$}
\State \textbf{inputs} $= Directory~D$
\State \textbf{node} $R$
\State $sum \gets 0$
\ForAll {$nodes~n \in D$}
\State $sum \gets sum+n.rank$
\EndFor
\State $x \gets Random[0,1)*sum$
\For {$nodes~i \leftarrow 0~to~D.length$}
\If {$x<D.element(i).rank$}
\State $R \gets D.element(i)$
\State \textbf{return} $R$
\Else
\State $x \gets x - D.element(i).rank$
\EndIf
\EndFor
\EndProcedure
\end{algorithmic}
\end{algorithm}		

In the next section, we evaluate the proposed Tor path selection technique.

\section{Evaluation}\label{eval}


There have been many different methodologies employed in the literature to evaluate proposals on the Tor anonymity network.
%
A number of proposals (e.g.~\cite{snader08,mccoy-pet2008,tissec-latency-leak}), have used the real-world Tor network for experimentation. Although, Soghoian~\cite{soghoian2012enforced} notes that carrying out experiments on the real Tor network has created problematic behaviors, for example affecting the overall quality of network or threatening the privacy of users. Moreover, in such experiment, the researchers are unable to modify the network and observe the effect on the Tor network and the anonymity it provides. On the other hand, a number of works (e.g.~\cite{wpes12-torchestra,acsac11-tortoise,pets2011-defenestrator} have employed experimenTor~\cite{cset11-experimentor}, with which one is able to emulate the Tor network while considering different network configurations.

%
%
%
%

As such, in this manuscript we employ the experimenTor \cite{cset11-experimentor} software as well, which is a large scale emulator that models a network topology using modelnet \cite{vahdat2002scalability}.
More specifically, we have used experimenTor on two separate machines that are connected with a gigabit Ethernet link. All of the Tor relays, directory servers, user nodes, and web pages run on the host machine which is a server with 12 core-2.4 GHz CPU and 18 Gigabytes of RAM, running Generic Ubuntu 3.2.0-27. The other machine is the emulator which runs a FreeBSD 6.3 OS, on a 2 core-2.0 GHz CPU with 2GB of RAM. Emulator has the role of the core of the network and all traffic between nodes passes through the emulator.

For modeling a topology like Internet topology on the emulator, we have used inet version 3 \cite{winick2002inet}. This tool models a network based on the way that Autonomous Systems are connected in the real Internet. We have simulated 50 web servers and 50 relays, where 5 of them are also authority servers, and a number of network infrastructure nodes such as routers, switches, and so on. For each link between the nodes, we have considered characteristics like bandwidth, delay, and loss.


\begin{figure}
\centering
\def\svgwidth{\columnwidth}
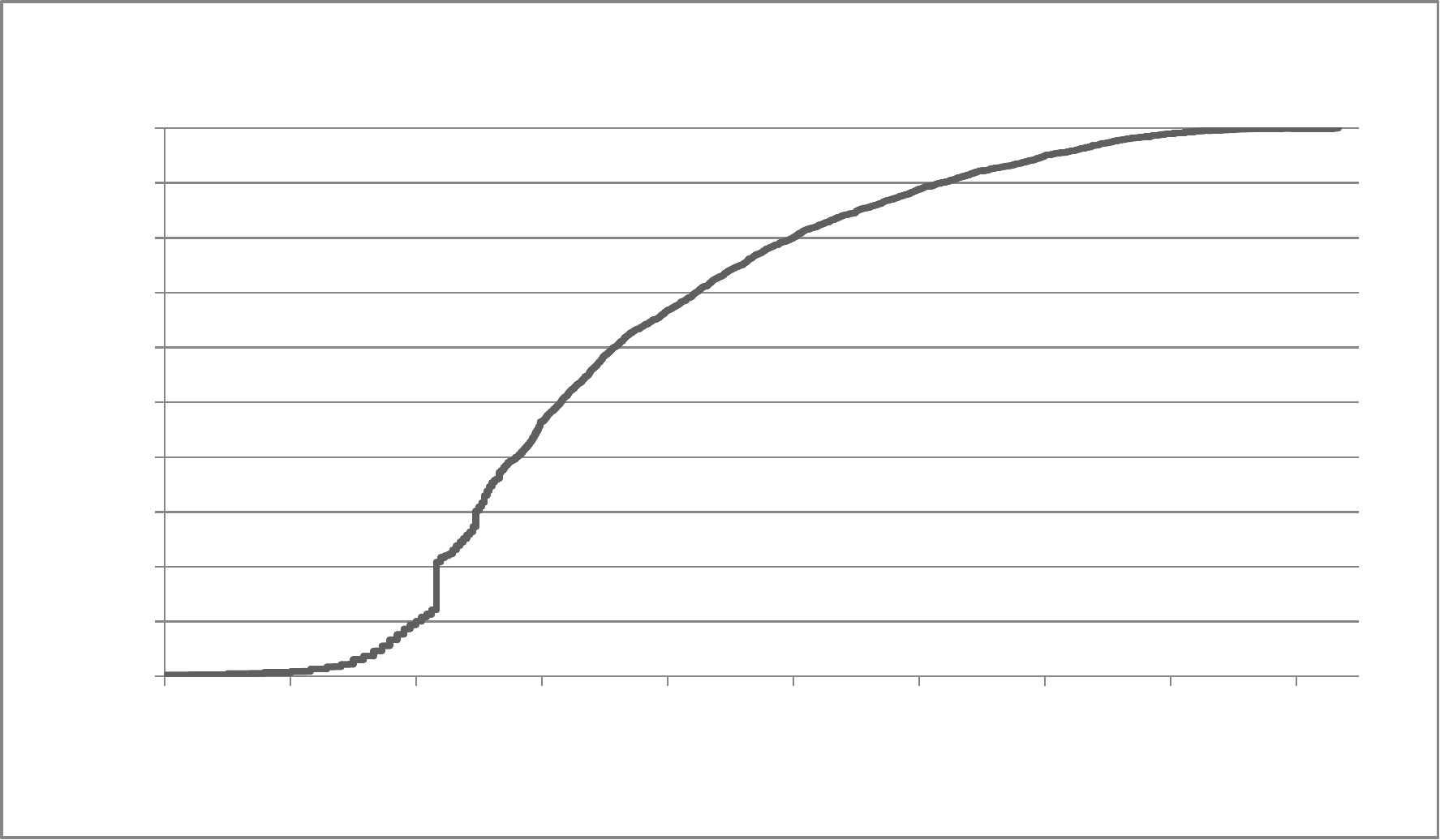
\caption{Cumulative distance of relay bandwidths in the Tor consensus-status document calculated by Tor Scanners in June 2012.}
\label{fig:relays}
\end{figure}
For assigning bandwidths of relays like the real Tor network, we have used the values which are calculated by the Tor scanners \cite{netscanners} and are recorded in the consensus-status document. The cumulative distribution of these values in the June 2012,which consists of 2966 relays, is shown in Figure \ref{fig:relays}. Approximately for each relay, the amount shown in this figure, is the minimum of its upload speed and download speed. We postulate that 10 percent of relays are institutional which have the same upload and download speeds and 90 percent of relays are residential which have download speed equal to twice of their upload speed. 

Afterwards we sample systematically from these relays with the sampling frame size of 60 and select 50 relays, where values from selected relays (i.e. rate-limiting bandwidth for normal and burst traffic, etc.). Furthermore, we have extracted the measured bandwidth for each sampled relay as calculated by Tor Scanners. As default path selection algorithm in Tor, selects relays based on that measured bandwidth and there is no network scanner in our simulation, we have hard-coded those bandwidth values in the source of their Tor software's code.

In this simulation, path selection and assigning streams to paths has been done through the control port. For this purpose we have implemented a java program based on JTorCtl \cite{JTorCtl} that creates a new path every ten minutes consisting of three relays.. With this method, we compare four different path selection methods with each other. For each path selection methodology the simulation was executed for 3 days. Below we will review the different path selection considered as part of our evaluation:

\begin{enumerate} 
\item {\bf Random:} Select 3 non-recurring relays randomly from the 50 available relays.
\item {\bf Default:} Employing the default path selection algorithm as used in Tor 0.2.2.35 (last available version when the simulations were conducted~\footnote{We have studied the change logs for the current Tor version  2.3.25 and to the best of our knowledge have found no updates to the path selection mechanism, since the Tor version we have used in our experiments.}), where relays are selected based on their bandwidth and uptime.
\item {\bf Geo:} Selecting path only based on minimizing latency to the average geo-graphical location as presented in Section \ref{geo}.  
\item {\bf Composite:} Using composite metric for path selection which has been mentioned in section \ref{geo}. This method benefits from both the advantages of second and third methods noted above.
\end{enumerate} 

We should note that as webservers are chosen randomly in our simulation, we consider no history in selecting the avg. geo-location point; so $3^{rd}$ relay will be selected randomly. This random selection will not affect the results positively, since avg. geo-location inclines the algorithm towards selecting circuits which end near to most visited websites by the user; but when no logic is behind choosing destination points, then the average geo-location point is selected randomly, this is equivalent  to selecting a random $3^{rd}$ relay.

Furthermore, as the emulated relays are not geographically distributed, before running the experiments we should create a map of Tor relays and specify the delay between them. So we measure RTT between relays by using ping packets, when no other traffic exists in the network and then given the observations made in~\cite{oakland2012-lastor} we postulate these RTT measurements as geographical distance between relays. We should note that as the experiment is being done and traffic packets are send over the simulated nodes, the RTT between relays will surely change, nevertheless we consider them as static given that these geographical locations will not change. More precisely we do not transmit any ping packets during the experiment and that is only done before the start of the evaluation phase.

\begin{figure}[h]
\centering
\def\svgwidth{\columnwidth}
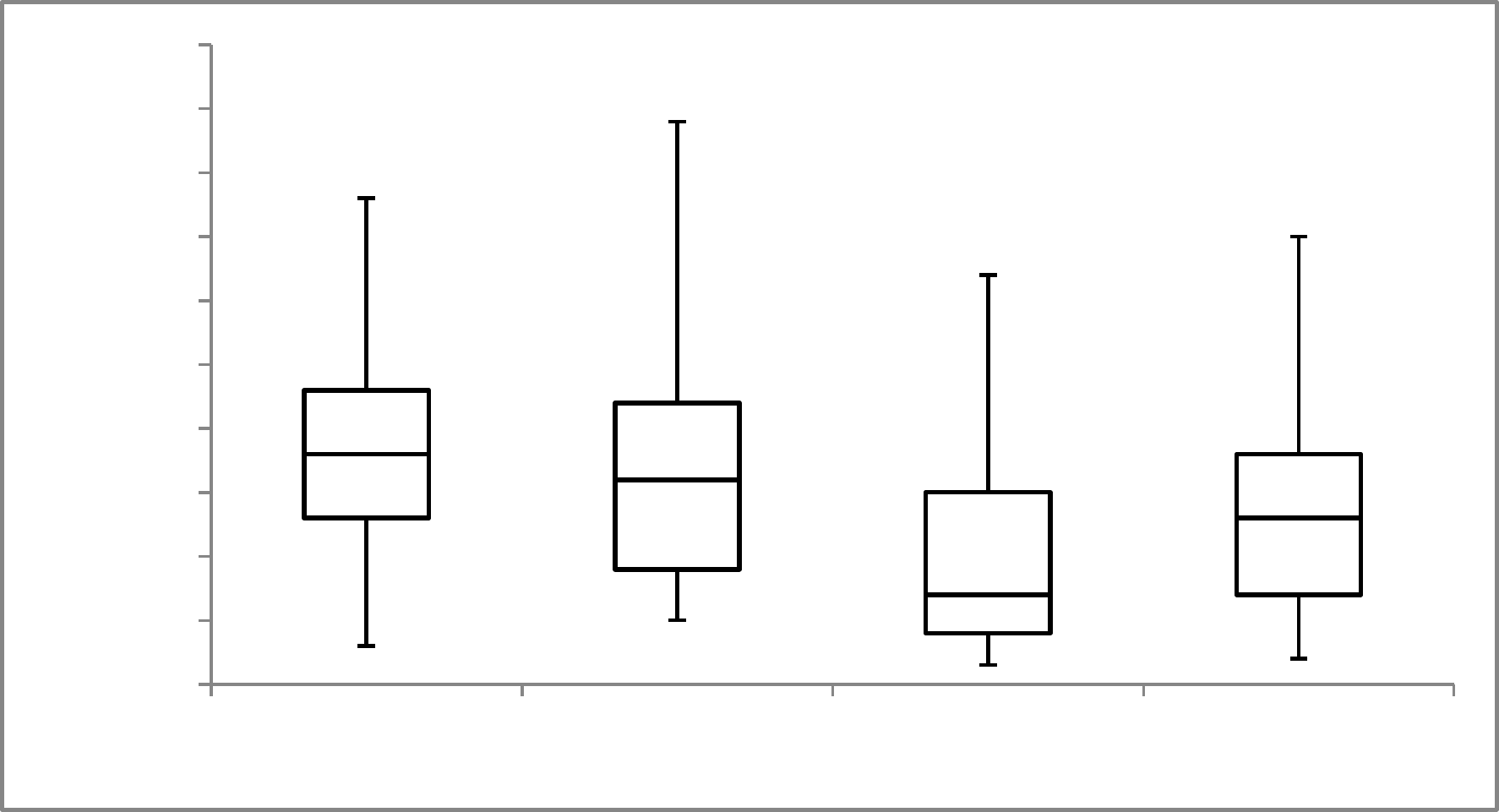
\caption{Time required for downloading 1-byte in different path selection methods.}
\label{fig:300_first}
\end{figure}

For measuring the performance of the simulated network, we consider two users. The first user is using Tor for web browsing and repeatedly downloads small size (i.e. 300 KB) web pages and the other user is using Tor for downloading large size files (i.e. 5MB) from random web servers. These behaviors are two different uses of Tor network which are most common between the users. We measure the time spent for each phase of downloading these files with a tool named torperf \cite{torperf}. 

Figure~\ref{fig:300_first} plots the time spent for receiving the first byte for each one of the methods. In this figure, least median is for the Geo method. Take note that as there are only two active users on the network there are no congested nodes, therefore when transferring only 1 byte of information the effect of relay bandwidths is minimal and propagation delay dominates.

\begin{figure}
\centering
\def\svgwidth{\columnwidth}
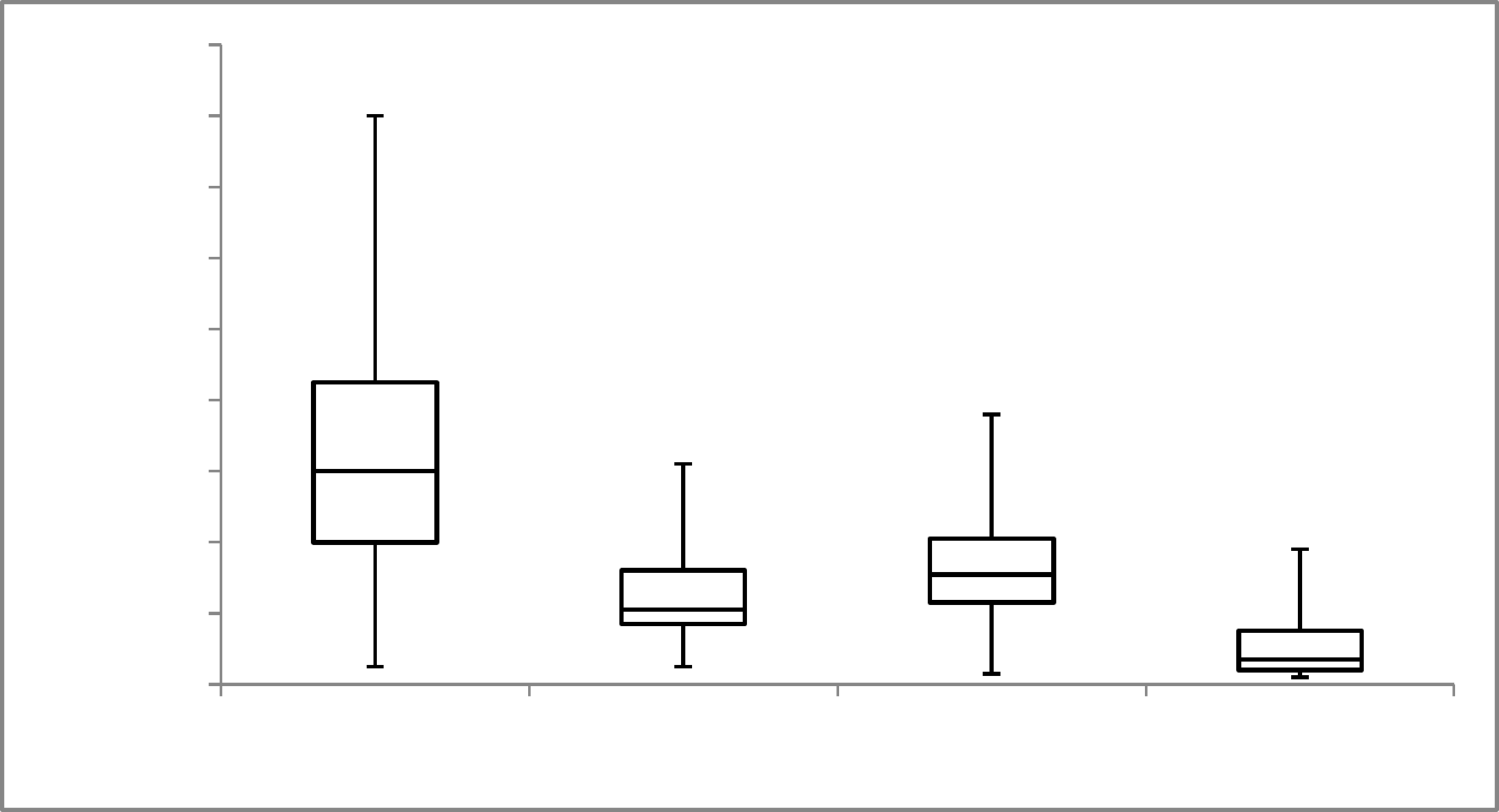
\caption{Time required for downloading 300 KB file in different path selection methods.}
\label{fig:300_dl}
\end{figure}
\begin{figure}
\centering
\def\svgwidth{\columnwidth}
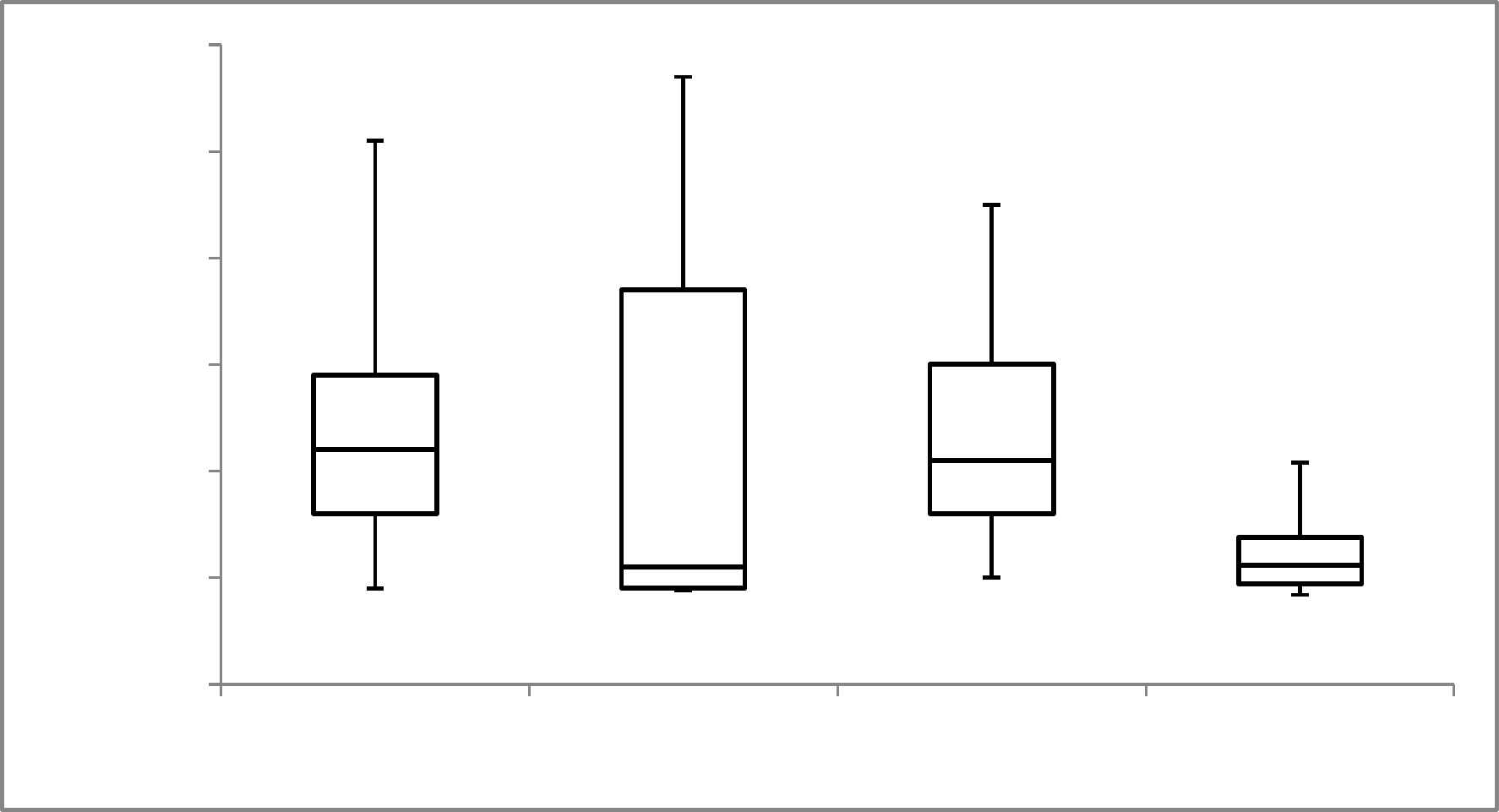
\caption{Time required for downloading 5 MB file in different path selection methods.}
\label{fig:5_dl}
\end{figure}

You can see the time spent for transferring 300 KB and 5 MB files in Figure~\ref{fig:300_dl} and \ref{fig:5_dl}. As observed, the default algorithm employed in Tor, performs better than the Geo algorithm. This is due to the fact that the Geo algorithm does not consider the bandwidth of relays when select a path. When transferring a small size packet (i.e. 1 byte) non of the relay's links will be congested; but when transferring larger size files, without considering bandwidth of relays, congestion could occur which results in the addition of queuing delay to the overall observed delay. Therefore, it is critical to additionally consider the bandwidth of relays. As observed in Figure~\ref{fig:300_dl}, combining methods that try to reduce the propagation delay while selecting higher bandwidth delays would be advantages.

Nevertheless methods that reduce propagation delay like Geo, work better for transferring small size files and methods that reduce queuing delay like default Tor, work better for transferring big files. So each method has a different performance with respect to the file size requested, therefore and for a better comparison, we should choose file sizes as the users really request in the real Tor network.

\subsection{Simulating normal behavior of users}
\begin{table}
\centering
\small
\begin{tabular}{|c|c|c|c|c|}
\hline
Country&Percentage&number&DL Speed&UL Speed\\
& & & (Mbps)& (Mbps)\\ \hline
US & $15 \%$ & 270 & 12.67 & 3.39\\ \hline
Germany & $10 \%$ & 180 & 14.67 & 2.14\\ \hline
Iran & $10 \%$ & 180 & 1.5 & 0.91\\ \hline
Italy & $10 \%$ & 180 & 5.46 & 1.09\\ \hline
France & $5 \%$ & 90 & 12.02 & 2.88\\ \hline
\end{tabular}
\caption{Top-5 Countries using Tor in Jun 2012 and modeling 900 clients based on their share and average bandwidth.}
\label{tab:users}
\end{table}
Here we model behavior of 900 clients. First we consider their bandwidths based on the real bandwidths of users in the real Tor network. According to reports from Tor metrics \cite{metrics} half of the users of Tor were from 5 countries in Jun 2012. We then assign average bandwidth of each country according to data on NetIndex \cite{netindex}, and partition 900 clients with respect to the percentages of users in those countries. (See Table~\ref{tab:users})

For simulating the behavior of clients, we partition them into two different groups. The first group which consists of 870 users, use Tor for web browsing. They request an HTML file and read that page in 5 to 15 seconds and then request another file and after visiting 150 pages they stop for a 15-20 minutes break. We simulate this behavior exactly based on \cite{hernandez2003tracking}. The second group, consisting of 30 users, are bulk downloaders which request large sized files one after another without interruption. According to \cite{mccoy-pet2008} thought they cover only 3\% of users but they makeup a large fraction of network traffic. 
\begin{figure}
\centering
\def\svgwidth{\columnwidth}
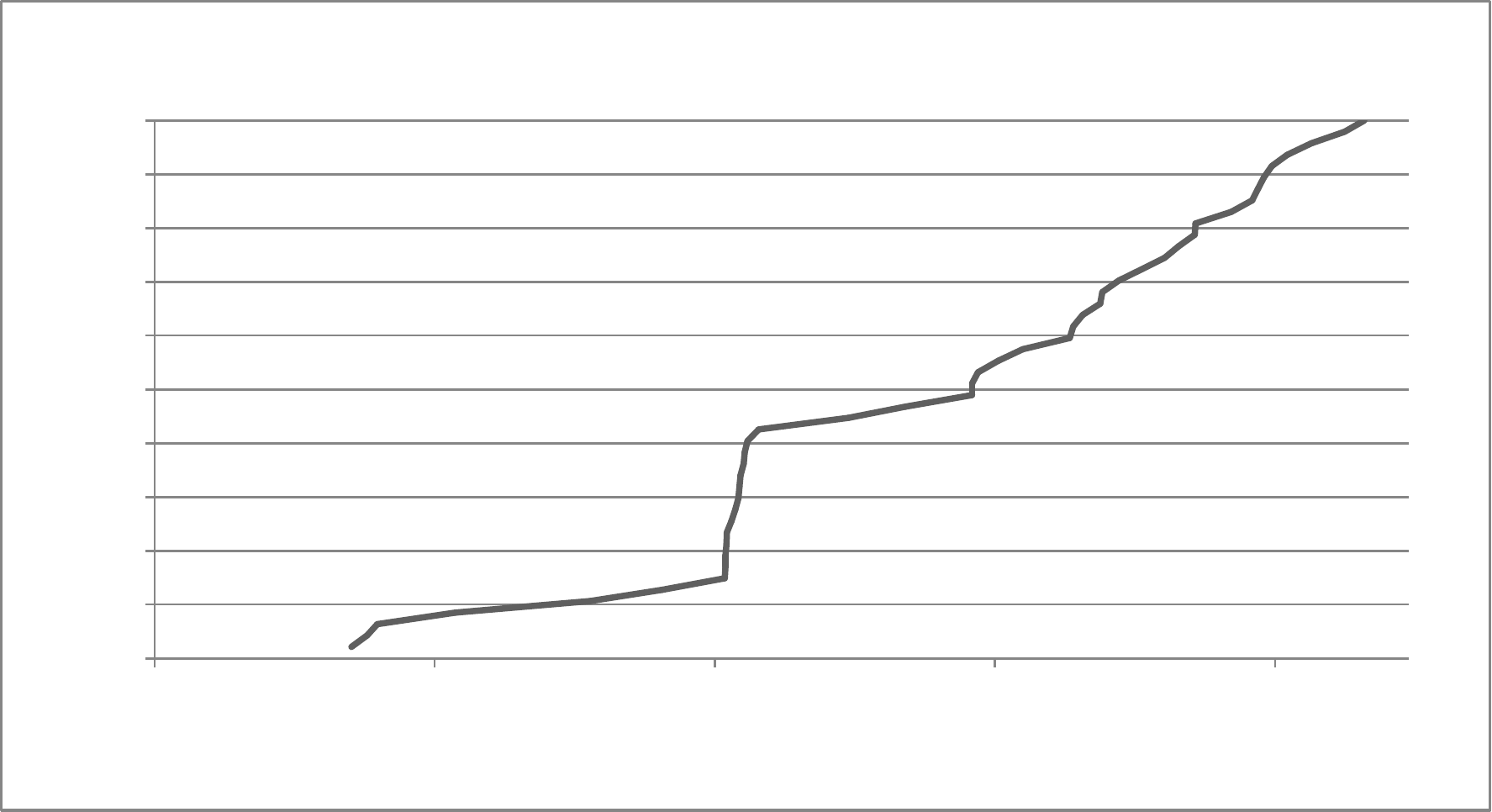
\caption{Cumulative distribution of front-page sizes of top 50 popular websites introduced by Alexa in Jun 2012.}
\label{fig:front}
\end{figure}

For measuring the average requested file sizes by normal users, we checked top 50 popular websites introduced by alexa \cite{alexa}. The cumulative fraction of them on Jun 2012 is available in Figure \ref{fig:front}. We selected the 10th, 30th, 50th, 70th, and 90th fractions which are equal to 3, 12, 82, 276, and 911 Kilobytes and intermittently each user requests one of them randomly. For bulk users, they request randomly one of the 1, 5, 10, 20, and 50 megabyte files. 
\begin{figure}
\centering
\def\svgwidth{\columnwidth}
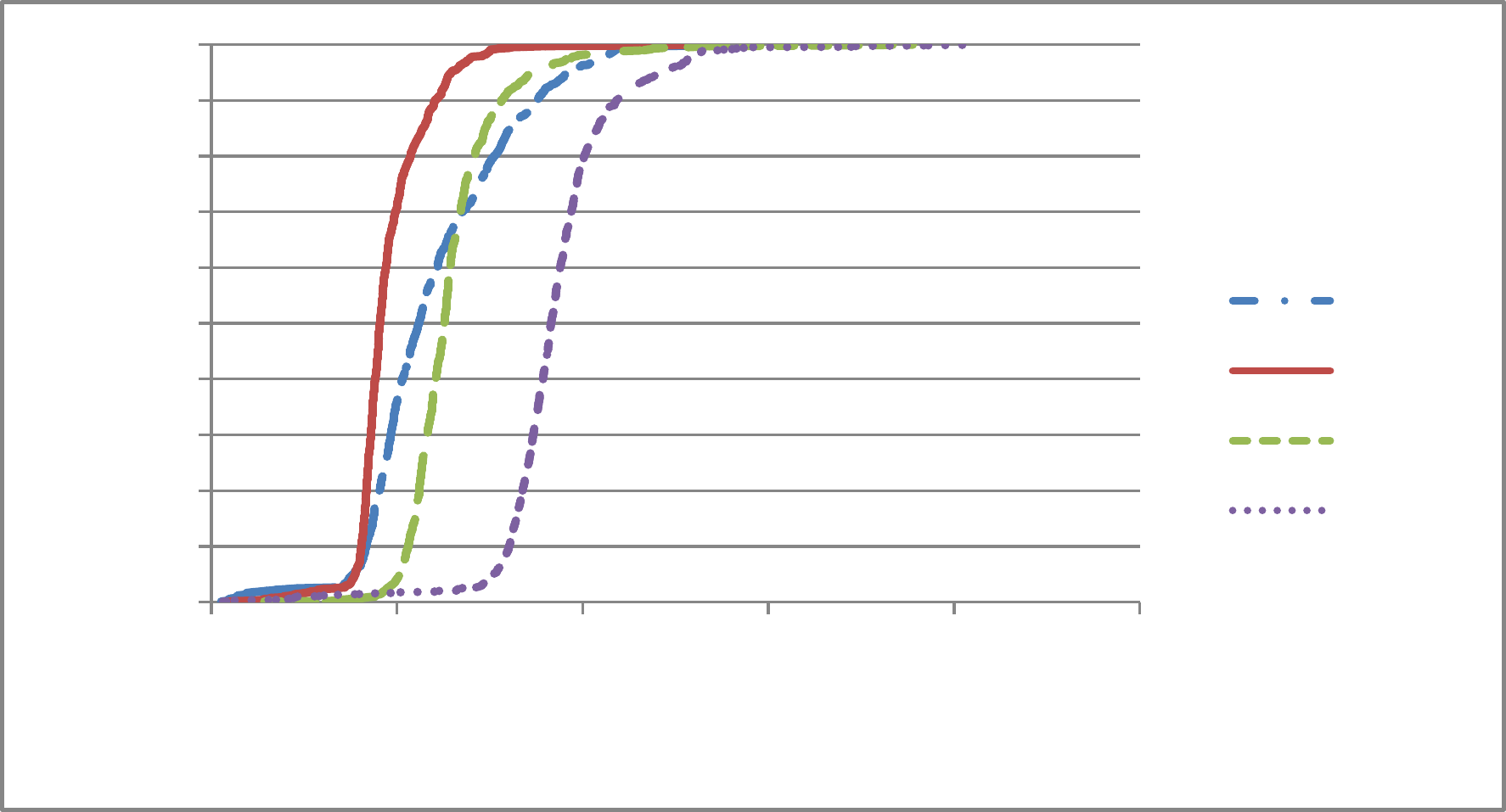
\caption{Cumulative distribution of average throughput of 900 clients in different path selection methods.}
\label{fig:wget}
\end{figure}

Figure~\ref{fig:wget} shows the impact of using different path selection methods on the performance. According to this figure, the bandwidth that user's experience in composite-metric method is approximately double that of the default Tor path selection. 
In the duration of the experiment for each path selection algorithm, which executed for about 3 days, approximately 135000 circuits were built. Each of the bulk downloaders, downloaded about 30 different files on average and each of the web browsing users requested about 1200 different web pages. For each request, we have calculated the throughput based on the time spent for the file transfer. Afterward we have calculated the average performance for each user by considering all throughput values obtained for that user.

Based on the Figure~\ref{fig:wget}, users using random path selection algorithm obtain the longest download times and they had the worst throughput. Comparing the Geo-algorithm with the default path selection algorithm of Tor, some users have experience better performance with our proposed Geo-algorithm and some obtain a better throughput with the default Tor algorithm. But Tor algorithm obtains a better median throughput.
%
Furthermore, the median throughput that user experience in composite-metric method is approximately 37KBps which is reasonably better than all other considered path selection techniques. 

It should be noted that in all the experiments, we have monitored the process and memory usage statistics of nodes to make sure that no resource issues influences the results. Also all of the four algorithms consume about the same resources and none would be considered preferable to the others in the context of resource usage.

\subsection{Analyzing anonymity}

Before using a new algorithm for path selection in Tor, analyzing its effect on the security of network is of great importance. One of the more prevalent attacks on Tor is the Sybil attack \cite{sybil}, in which the attacker takes control of a fraction of network relays. If the first and the last selected relays by the user are under control of the attacker then he/she can violate the anonymity by carrying out a correlation attack. In what follows we will employ the methodology used in~\cite{panchenko2009path} to analyze the security of the proposed technique.
\begin{table}
\centering
\small
\begin{tabular}{|c|c|c|c|c|c|c|}
\hline
&first&middle&end&start-end&E(x)&d\\
& & & &comb& &\\ \hline
Random & 50 & 50 & 50 & 2450 & 11.177 & 0.990\\ \hline
default & 19 & 45 & 45 & 625 & 7.277 & 0.644\\ \hline
Geo & 50 & 50 & 50 & 2450 & 11.155 & 0.988\\ \hline
Composite & 50 & 50 & 50 & 1947 & 8.953 & 0.793\\ \hline
\end{tabular}
\caption{Number of distinct relays selected for different path selection methods and their entropy and anonymity degree.}
\label{tab:ee}
\end{table}

Based on~\cite{Diaz02} and as employed in~~\cite{panchenko2009path}, the anonymity degree is defined as 
$d = \dfrac{E(X)}{E_{max}} $
which E(x) is the entropy and is computed through this relation:
\begin{eqnarray*}
\label{eqn:antropy}
E(X) = -\sum\limits_{i=1}^{N} p_i\log_2(p_i)
\end{eqnarray*}
In the above relation N is the number of all different start-end combinations and $p_{i}$ is the probability of selection of that combination. So in the best situation, maximum entropy is when every combination has an equal probability of 
$\dfrac{1}{N}$
for selection.

For comparing different path selection methods we should calculate the probability with which the first and last selected relays are under the control of an attacker. In the duration of our simulation, each of the 900 clients creates about 500 circuits. Approximately 150 of the 500 circuits are used and the rest are created as backup circuits. This is do to the fact that if one of the relays used in an active circuit is suddenly lost, then the clients should not restart the circuit creation process all over as it would delay the communication. Therefore, each client creates 2 or 3 backup circuits, which is not employed unless the primary circuit fails. 

For each path selection method, we have analyzed about 135,000 start-end combinations for calculating anonymity degree of each path selection method. As we have used 50 relays in our evaluation, the maximum number of different start-end combinations will be 50*50=2500. In the best situation, a path selection method chooses all the start-end combinations with equal probability where this probability is 1/2500. So the maximum entropy is:

\begin{eqnarray*}
\label{eqn:antropy}
H_{max} = -\sum\limits_{i=1}^{2500} (1/2500) \log_2(1/2500) = 11.28
\end{eqnarray*}

After analyzing used relays in paths selected by each method, the number of different start-end combinations has been measured and is shown in Table~\ref{tab:ee}, where the calculated entropy and anonymity degree for each path selection method is also presented.

As observed, the default path selection has the lowest anonymity degree and in turn the worst security. In RTT method relays are selected based on the location of sender and receiver and because of their diversity, different start-end combinations have been chosen and anonymity degree is as high as random path selection. Anonymity degree of composite-metric path selection algorithm is about 79\%  which is better than default Tor path selection. As illustrated earlier, the composite-metric path selection also has better performance than the default Tor path selection, therefore we believe that it is a good replacement for the current path selection algorithms.

\section{discussion}\label{dis}

In this section we analyze some key design characteristics of anonymous networks which are of great importance. We compare our proposed method to some related works in the context of each characteristic.

{\bf Low Computational Cost:}
There are two approaches for selecting circuits in an anonymous network. One approach is selecting nodes one after another and the other one is selecting a complete circuit at once. The important point is the processing overhead incurred when selecting nodes one after another as compared to when selecting a circuit containing three nodes. Assume that there are n relays available in the Tor network; if one wants to select nodes one after another, he should process n nodes three times for making a circuit i.e. O(n). But if one wants to select a path containing three nodes, he should process all possible paths which is about $O(n^{3})$. 

Given the growing number of Tor relays, latency based methods which analyze all possible circuits require large processing times. 
Many latency-based path selection methods in the literature (i.e.  \cite{DBLP:conf/pet/SherrBL09,panchenko2009path,oakland2012-lastor}) select a complete circuit containing three nodes. For decreasing processing time our proposed algorithm selects nodes one after another instead of selecting the whole circuit, hence it incurs a much smaller processing overhead.

\textbf {Unpredictable Path:} Using Average-Geo-Location point brings a lot of advantages. First it makes our algorithm independent of destination IP address. In the method proposed in \cite{oakland2012-lastor} which is dependent to destination IP address, users should make a new circuit for each destination they want to make a connection with. Where making a new circuit for each destination the client wants to visit, would lengthen the time it takes for the client to receive the content of interest. In contrast, our algorithm can make an optimal circuit and use that circuit for many destination IP addresses because the circuit is selected based on the user behavior and it is an average point of possible destination addresses used by that specific user.

Moreover, using the average geo-location destination makes network more invulnerable to attacks. In latency based methods if an attacker wants to prove a specific source connection with a destination nearby, he/she can run many relays between the geographical path along the source and destination. Then the probability that the user selects a malicious relay increases. But with average geo-location the attacker must do more and determine the $3^{rd}$ relay position based on the user's usage history. As its difficult for an attacker to know the usage history of a user, and the fact that the position of average-geo-location point changes over time, it is almost impossible for an attacker to guess the path and run relays in that path.

%

\textbf{Low Traffic Overhead:} Our path selection method uses a locally available IP-geo-location file for measuring coordinates of relays on a map. So the path selection process could be done passively without generating any extra traffic. Nevertheless \cite{DBLP:conf/pet/SherrBL09,panchenko2009path} should generate a lot of traffic for selecting a proper path. In the former one, each user would send many ping packets for calculating distances and a lot of traffic would be required for transferring coordinate system information. Also in the latter some traffic would be generated for violating exit policy, with which feedback is provided to the sender for estimating the path RTT.

\textbf{Ease of Implementation:}
Besides Using our path selection method only needs local modifications on the client side and it is possible for some users to employ the proposed method in collaboration with other releases of Tor. In contrast, techniques such as~\cite{DBLP:conf/pet/SherrBL09} are not practically usable i.e. all relays should participate in constructing a coordinate system, that would require that all relays, directory servers, and clients be upgraded.

\textbf{Minimal Setup Time:} As soon as clients start the Tor browser, they can optimally select circuits in our path selection mechanism but in \cite{panchenko2009path,congestion-tor12} some time will be required so that users can calculate latency between an appropriate number of relays, and in turn select an optimal path.

As there is no practical implementation available for most of the related works, we could not evaluate the performance and anonymity degree provided by these works in the same test environment. Nevertheless, we provide comparison based on the points discussed above in In Table~\ref{tab:com}. 

There are two important issues which we should address in the remainder of this section. First and in order to prevent the predecessor attack, Wright et al.~\cite{wright03} proposed ``Entry Guards". In the current Tor network, three Guard nodes are randomly selected from relays with high bandwidth and high uptime values. In earlier latency-based path selection methods, Akhoondi  et al.~\cite{oakland2012-lastor} propose the selection of three nearby nodes in three different geographical directions for resisting such attacks. In our proposed composite-metric path selection, we can combine the two noted approaches and select three nearby relays with high bandwidth and uptime values, as entry guards, which are located in three different geographical directions. Hence, an adversary would be prevented from conducting a predecessor attack on the network.

Second and more importantly, one may argue that if the user only visits a specific set of websites, then an attacker could guess the average geo-location point and mount an attack on the Tor network. Such complications are inevitable when employing any logical path selection algorithms, as we are preferring some relays over others. The most secure path selection method in the Tor network, is random relay selection; with which there is no preferred relay among all others. In order to provide better performance, different path selection algorithm identify a set of  relays as more preferable to others based on some logic.  

For instance, in path selection methods which work based on bandwidth of relays (i.e. Tor’s default algorithm), high-bandwidth relays have a higher chance of being selected by the users. Alternatively, in latency-based path selection methods (i.e. considering geographical locations), an adversary who knows a user often connects to a specific server, can set up controlled relays somewhere in the path between user and the destination server, although this would be much harder than just setting up a high bandwidth relay.

In the proposed method, an attacker should first obtain the position of the average geo-location point and then setup relays in that area to improve his/her chances of being selected by the client. This would be quite hard, as the average geo-location point changes over time and is not fixed, furthermore the attacker would be required to setup high-bandwidth relays with high up-time values which are close to a changing average geo-location point.

The only exception is when the client is only using the Tor network to connect to a single website every time, then as the average geo-location would be constant, he/she would limit the random selection of the Tor relays to a smaller set of possible relays. This is due to the fact that the proposed relay selection methodology is based on random selection from relays ranked based on multiple parameters (i.e. geographical distance, bandwidth, and up-time). Therefore, even though the avg. geo-location could be fixed, the other two parameters still affect the relay selection and we are not limited to a single relay. We do believe that such usage scenario, where only a single specific website is visited, is not observed widely. Nevertheless, in such scenario the attacker would have to run a high bandwidth, and high up-time relay near the avg. geo-location. That is not at all easy, nor straight forward, specially when compared to the current Tor implementation in which an attacker just needs to run a high bandwidth and high up-time server anywhere on the Internet in order to attract traffic from the users on the Tor network
%


We believe that given the comprehensive evaluation conducted in this work, and as illustrated by the obtained results, the proposed technique has better performance and security in comparison with current Tor default path selection algorithm; while it has benefits which are not available in previously proposed techniques.

\begin{table}
\centering
\small
\begin{tabular}{|c|c|c|c|c|c|}
\hline
 & \cite{DBLP:conf/pet/SherrBL09} & \cite{panchenko2009path} & \cite{congestion-tor12} & \cite{oakland2012-lastor} & Our Work \\ \hline 
Low Computational Cost & \xmark & \xmark & \cmark & \xmark & \cmark \\ \hline
Unpredictable Path & \cmark & \cmark & \cmark & \xmark & \cmark \\ \hline
Low Traffic Overhead & \xmark & \xmark & \cmark & \cmark & \cmark \\ \hline
Ease of Implementation & \xmark & \cmark & \cmark & \cmark & \cmark \\ \hline
Minimal Setup Time & \cmark & \xmark & \xmark & \cmark & \cmark \\ \hline
\end{tabular}
\caption{Comparison with related works}
\label{tab:com}
\end{table}

\section{Conclusion and future works}\label{con}
In this paper, we proposed an algorithm for improving performance and security of the Tor network, by employing a combination of different metrics in the process of path selection between the source and destination nodes. These metrics are bandwidth and uptime of relays as node conditions and geographical locations of relays as path conditions.  Using our novel solution, we could double the  performance and in addition, our method has greater anonymity degree than the default path selection algorithm of Tor. 

As part of the future works, we are currently pursuing the deployment of the proposed path selection methodology on the PlanetLab~\cite{peterson2006planetlab} platform for a more accurate evaluation, specially when it comes to selecting entry guards on three different geographical directions as noted in the previous section. 
Furthermore, we believe that by considering more characteristics like congestion, load, and so on, it is possible to achieve better performance and security in the future.

%
%
%
%



\section*{Acknowledgment}
The authors would like to thank the anonymous reviewers for their constructive and detailed comments, with which we were able to improve this manuscript.  This work was partially supported by ITRC under grant number 17179/500 (90/11/30).

\bibliographystyle{IEEEbib}
\bibliography{test}

\end{document}

%% file: img/relays.pdf_tex
\begingroup%
\begin{tiny}
  \makeatletter%
  \providecommand\color[2][]{%
    \errmessage{(Inkscape) Color is used for the text in Inkscape, but the package 'color.sty' is not loaded}%
    \renewcommand\color[2][]{}%
  }%
  \providecommand\transparent[1]{%
    \errmessage{(Inkscape) Transparency is used (non-zero) for the text in Inkscape, but the package 'transparent.sty' is not loaded}%
    \renewcommand\transparent[1]{}%
  }%
  \providecommand\rotatebox[2]{#2}%
  \ifx\svgwidth\undefined%
    \setlength{\unitlength}{511.0000009bp}%
    \ifx\svgscale\undefined%
      \relax%
    \else%
      \setlength{\unitlength}{\unitlength * \real{\svgscale}}%
    \fi%
  \else%
    \setlength{\unitlength}{\svgwidth}%
  \fi%
  \global\let\svgwidth\undefined%
  \global\let\svgscale\undefined%
  \makeatother%
  \begin{picture}(1,0.58341666)%
    \put(0,0){\includegraphics[width=\unitlength]{relays.pdf}}%
    \put(0.08418934,0.10726696){\makebox(0,0)[lb]{\smash{0}}}%
    \put(0.06845548,0.14536872){\makebox(0,0)[lb]{\smash{0.1}}}%
    \put(0.06845548,0.18349005){\makebox(0,0)[lb]{\smash{0.2}}}%
    \put(0.06845548,0.22159181){\makebox(0,0)[lb]{\smash{0.3}}}%
    \put(0.06845548,0.25973271){\makebox(0,0)[lb]{\smash{0.4}}}%
    \put(0.06845548,0.2978149){\makebox(0,0)[lb]{\smash{0.5}}}%
    \put(0.06845548,0.3359558){\makebox(0,0)[lb]{\smash{0.6}}}%
    \put(0.06845548,0.37403799){\makebox(0,0)[lb]{\smash{0.7}}}%
    \put(0.06845548,0.41217889){\makebox(0,0)[lb]{\smash{0.8}}}%
    \put(0.06845548,0.45026108){\makebox(0,0)[lb]{\smash{0.9}}}%
    \put(0.08418934,0.48840198){\makebox(0,0)[lb]{\smash{1}}}%
    \put(0.10890166,0.07397928){\makebox(0,0)[lb]{\smash{1}}}%
    \put(0.19625979,0.07397928){\makebox(0,0)[lb]{\smash{4}}}%
    \put(0.27835372,0.07397928){\makebox(0,0)[lb]{\smash{16}}}%
    \put(0.36575098,0.07397928){\makebox(0,0)[lb]{\smash{64}}}%
    \put(0.44784491,0.07397928){\makebox(0,0)[lb]{\smash{256}}}%
    \put(0.52995842,0.07397928){\makebox(0,0)[lb]{\smash{1024}}}%
    \put(0.61731654,0.07397928){\makebox(0,0)[lb]{\smash{4096}}}%
    \put(0.69941047,0.07397928){\makebox(0,0)[lb]{\smash{16384}}}%
    \put(0.78676859,0.07397928){\makebox(0,0)[lb]{\smash{65536}}}%
    \put(0.86890166,0.07397928){\makebox(0,0)[lb]{\smash{262144}}}%
    \put(0.04703865,0.21646461){\rotatebox{90}{\makebox(0,0)[lb]{\smash{Cumulative Fraction}}}}%
    \put(0.45109345,0.03272684){\makebox(0,0)[lb]{\smash{BandWidth (KB/s)}}}%
    \put(0.35739482,0.53073075){\makebox(0,0)[lb]{\smash{Relay BandWidths}}}%
  \end{picture}%
\end{tiny}
\endgroup%

%% file: img/first_byte.pdf_tex
\begingroup%
\begin{tiny}
  \makeatletter%
  \providecommand\color[2][]{%
    \errmessage{(Inkscape) Color is used for the text in Inkscape, but the package 'color.sty' is not loaded}%
    \renewcommand\color[2][]{}%
  }%
  \providecommand\transparent[1]{%
    \errmessage{(Inkscape) Transparency is used (non-zero) for the text in Inkscape, but the package 'transparent.sty' is not loaded}%
    \renewcommand\transparent[1]{}%
  }%
  \providecommand\rotatebox[2]{#2}%
  \ifx\svgwidth\undefined%
    \setlength{\unitlength}{510.78133119bp}%
    \ifx\svgscale\undefined%
      \relax%
    \else%
      \setlength{\unitlength}{\unitlength * \real{\svgscale}}%
    \fi%
  \else%
    \setlength{\unitlength}{\svgwidth}%
  \fi%
  \global\let\svgwidth\undefined%
  \global\let\svgscale\undefined%
  \makeatother%
  \begin{picture}(1,0.54163926)%
    \put(0,0){\includegraphics[width=\unitlength]{first_byte.pdf}}%
    \put(0.1037701,0.0775855){\makebox(0,0)[lb]{\smash{0}}}%
    \put(0.08440369,0.12024563){\makebox(0,0)[lb]{\smash{0.5}}}%
    \put(0.1037701,0.16290577){\makebox(0,0)[lb]{\smash{1}}}%
    \put(0.08440369,0.20562463){\makebox(0,0)[lb]{\smash{1.5}}}%
    \put(0.1037701,0.24828477){\makebox(0,0)[lb]{\smash{2}}}%
    \put(0.08440369,0.2909449){\makebox(0,0)[lb]{\smash{2.5}}}%
    \put(0.1037701,0.33366377){\makebox(0,0)[lb]{\smash{3}}}%
    \put(0.08440369,0.3763239){\makebox(0,0)[lb]{\smash{3.5}}}%
    \put(0.1037701,0.41898404){\makebox(0,0)[lb]{\smash{4}}}%
    \put(0.08440369,0.4617029){\makebox(0,0)[lb]{\smash{4.5}}}%
    \put(0.1037701,0.50436304){\makebox(0,0)[lb]{\smash{5}}}%
    \put(0.20060215,0.03651117){\makebox(0,0)[lb]{\smash{Random}}}%
    \put(0.41358957,0.03651117){\makebox(0,0)[lb]{\smash{Default}}}%
    \put(0.63801046,0.03651117){\makebox(0,0)[lb]{\smash{Geo}}}%
    \put(0.8107654,0.03651117){\makebox(0,0)[lb]{\smash{Composite}}}%
    \put(0.0579501,0.25772129){\rotatebox{90}{\makebox(0,0)[lb]{\smash{time (s)}}}}%
  \end{picture}%
\end{tiny}
\endgroup%

%% file: img/300KB.pdf_tex
\begingroup%
\begin{tiny}
  \makeatletter%
  \providecommand\color[2][]{%
    \errmessage{(Inkscape) Color is used for the text in Inkscape, but the package 'color.sty' is not loaded}%
    \renewcommand\color[2][]{}%
  }%
  \providecommand\transparent[1]{%
    \errmessage{(Inkscape) Transparency is used (non-zero) for the text in Inkscape, but the package 'transparent.sty' is not loaded}%
    \renewcommand\transparent[1]{}%
  }%
  \providecommand\rotatebox[2]{#2}%
  \ifx\svgwidth\undefined%
    \setlength{\unitlength}{510.78133119bp}%
    \ifx\svgscale\undefined%
      \relax%
    \else%
      \setlength{\unitlength}{\unitlength * \real{\svgscale}}%
    \fi%
  \else%
    \setlength{\unitlength}{\svgwidth}%
  \fi%
  \global\let\svgwidth\undefined%
  \global\let\svgscale\undefined%
  \makeatother%
  \begin{picture}(1,0.54163926)%
    \put(0,0){\includegraphics[width=\unitlength]{300KB.pdf}}%
    \put(0.11034826,0.0775855){\makebox(0,0)[lb]{\smash{0}}}%
    \put(0.09732899,0.12498347){\makebox(0,0)[lb]{\smash{20}}}%
    \put(0.09732899,0.17240102){\makebox(0,0)[lb]{\smash{40}}}%
    \put(0.09732899,0.21985773){\makebox(0,0)[lb]{\smash{60}}}%
    \put(0.09732899,0.2672557){\makebox(0,0)[lb]{\smash{80}}}%
    \put(0.08426273,0.31467325){\makebox(0,0)[lb]{\smash{100}}}%
    \put(0.08426273,0.36209081){\makebox(0,0)[lb]{\smash{120}}}%
    \put(0.08426273,0.40954751){\makebox(0,0)[lb]{\smash{140}}}%
    \put(0.08426273,0.45694549){\makebox(0,0)[lb]{\smash{160}}}%
    \put(0.08426273,0.50436304){\makebox(0,0)[lb]{\smash{180}}}%
    \put(0.20657339,0.03651117){\makebox(0,0)[lb]{\smash{Random}}}%
    \put(0.41758345,0.03651117){\makebox(0,0)[lb]{\smash{Default}}}%
    \put(0.64067305,0.03651117){\makebox(0,0)[lb]{\smash{Geo}}}%
    \put(0.81152894,0.03651117){\makebox(0,0)[lb]{\smash{Composite}}}%
    \put(0.0579501,0.25772129){\rotatebox{90}{\makebox(0,0)[lb]{\smash{time (s)}}}}%
  \end{picture}%
\end{tiny}
\endgroup%

%% file: img/5MB.pdf_tex
\begingroup%
\begin{tiny}
  \makeatletter%
  \providecommand\color[2][]{%
    \errmessage{(Inkscape) Color is used for the text in Inkscape, but the package 'color.sty' is not loaded}%
    \renewcommand\color[2][]{}%
  }%
  \providecommand\transparent[1]{%
    \errmessage{(Inkscape) Transparency is used (non-zero) for the text in Inkscape, but the package 'transparent.sty' is not loaded}%
    \renewcommand\transparent[1]{}%
  }%
  \providecommand\rotatebox[2]{#2}%
  \ifx\svgwidth\undefined%
    \setlength{\unitlength}{510.78133119bp}%
    \ifx\svgscale\undefined%
      \relax%
    \else%
      \setlength{\unitlength}{\unitlength * \real{\svgscale}}%
    \fi%
  \else%
    \setlength{\unitlength}{\svgwidth}%
  \fi%
  \global\let\svgwidth\undefined%
  \global\let\svgscale\undefined%
  \makeatother%
  \begin{picture}(1,0.54163926)%
    \put(0,0){\includegraphics[width=\unitlength]{5MB.pdf}}%
    \put(0.11034826,0.0775855){\makebox(0,0)[lb]{\smash{0}}}%
    \put(0.09732899,0.14871182){\makebox(0,0)[lb]{\smash{50}}}%
    \put(0.08426273,0.21985773){\makebox(0,0)[lb]{\smash{100}}}%
    \put(0.08426273,0.2909449){\makebox(0,0)[lb]{\smash{150}}}%
    \put(0.08426273,0.36209081){\makebox(0,0)[lb]{\smash{200}}}%
    \put(0.08426273,0.43321713){\makebox(0,0)[lb]{\smash{250}}}%
    \put(0.08426273,0.50436304){\makebox(0,0)[lb]{\smash{300}}}%
    \put(0.20657339,0.03651117){\makebox(0,0)[lb]{\smash{Random}}}%
    \put(0.41758345,0.03651117){\makebox(0,0)[lb]{\smash{Default}}}%
    \put(0.64067305,0.03651117){\makebox(0,0)[lb]{\smash{Geo}}}%
    \put(0.81152894,0.03651117){\makebox(0,0)[lb]{\smash{Composite}}}%
    \put(0.0579501,0.25772129){\rotatebox{90}{\makebox(0,0)[lb]{\smash{time (s)}}}}%
  \end{picture}%
\end{tiny}
\endgroup%

%% file: img/frontpage.pdf_tex
\begingroup%
\begin{tiny}
  \makeatletter%
  \providecommand\color[2][]{%
    \errmessage{(Inkscape) Color is used for the text in Inkscape, but the package 'color.sty' is not loaded}%
    \renewcommand\color[2][]{}%
  }%
  \providecommand\transparent[1]{%
    \errmessage{(Inkscape) Transparency is used (non-zero) for the text in Inkscape, but the package 'transparent.sty' is not loaded}%
    \renewcommand\transparent[1]{}%
  }%
  \providecommand\rotatebox[2]{#2}%
  \ifx\svgwidth\undefined%
    \setlength{\unitlength}{511.00002748bp}%
    \ifx\svgscale\undefined%
      \relax%
    \else%
      \setlength{\unitlength}{\unitlength * \real{\svgscale}}%
    \fi%
  \else%
    \setlength{\unitlength}{\svgwidth}%
  \fi%
  \global\let\svgwidth\undefined%
  \global\let\svgscale\undefined%
  \makeatother%
  \begin{picture}(1,0.54416151)%
    \put(0,0){\includegraphics[width=\unitlength]{frontpage.pdf}}%
    \put(0.07631756,0.09721682){\makebox(0,0)[lb]{\smash{0}}}%
    \put(0.06208665,0.13332249){\makebox(0,0)[lb]{\smash{0.1}}}%
    \put(0.06208665,0.16938903){\makebox(0,0)[lb]{\smash{0.2}}}%
    \put(0.06208665,0.20547513){\makebox(0,0)[lb]{\smash{0.3}}}%
    \put(0.06208665,0.24154167){\makebox(0,0)[lb]{\smash{0.4}}}%
    \put(0.06208665,0.2776082){\makebox(0,0)[lb]{\smash{0.5}}}%
    \put(0.06208665,0.31367474){\makebox(0,0)[lb]{\smash{0.6}}}%
    \put(0.06208665,0.34979998){\makebox(0,0)[lb]{\smash{0.7}}}%
    \put(0.06208665,0.38586651){\makebox(0,0)[lb]{\smash{0.8}}}%
    \put(0.06208665,0.42193305){\makebox(0,0)[lb]{\smash{0.9}}}%
    \put(0.07631756,0.45799958){\makebox(0,0)[lb]{\smash{1}}}%
    \put(0.08918645,0.06700156){\makebox(0,0)[lb]{\smash{100}}}%
    \put(0.27226276,0.06700156){\makebox(0,0)[lb]{\smash{1000}}}%
    \put(0.45539386,0.06700156){\makebox(0,0)[lb]{\smash{10000}}}%
    \put(0.63846626,0.06700156){\makebox(0,0)[lb]{\smash{100000}}}%
    \put(0.8215778,0.06700156){\makebox(0,0)[lb]{\smash{1000000}}}%
    \put(0.04264242,0.20390957){\rotatebox{90}{\makebox(0,0)[lb]{\smash{Cumulative Fraction}}}}%
    \put(0.48304553,0.02966301){\makebox(0,0)[lb]{\smash{Size (Byte)}}}%
    \put(0.39551128,0.49637531){\makebox(0,0)[lb]{\smash{front}}}%
    \put(0.46623535,0.49637531){\makebox(0,0)[lb]{\smash{-}}}%
    \put(0.47670502,0.49637531){\makebox(0,0)[lb]{\smash{page size}}}%
  \end{picture}%
\end{tiny}
\endgroup%

%% file: img/wget.pdf_tex
\begingroup%
\begin{tiny}
  \makeatletter%
  \providecommand\color[2][]{%
    \errmessage{(Inkscape) Color is used for the text in Inkscape, but the package 'color.sty' is not loaded}%
    \renewcommand\color[2][]{}%
  }%
  \providecommand\transparent[1]{%
    \errmessage{(Inkscape) Transparency is used (non-zero) for the text in Inkscape, but the package 'transparent.sty' is not loaded}%
    \renewcommand\transparent[1]{}%
  }%
  \providecommand\rotatebox[2]{#2}%
  \ifx\svgwidth\undefined%
    \setlength{\unitlength}{510.87606888bp}%
    \ifx\svgscale\undefined%
      \relax%
    \else%
      \setlength{\unitlength}{\unitlength * \real{\svgscale}}%
    \fi%
  \else%
    \setlength{\unitlength}{\svgwidth}%
  \fi%
  \global\let\svgwidth\undefined%
  \global\let\svgscale\undefined%
  \makeatother%
  \begin{picture}(1,0.53922654)%
    \put(0,0){\includegraphics[width=\unitlength]{wget.pdf}}%
    \put(0.10358152,0.13173534){\makebox(0,0)[lb]{\smash{0}}}%
    \put(0.08399555,0.16875019){\makebox(0,0)[lb]{\smash{0.1}}}%
    \put(0.08399555,0.20578461){\makebox(0,0)[lb]{\smash{0.2}}}%
    \put(0.08399555,0.24283861){\makebox(0,0)[lb]{\smash{0.3}}}%
    \put(0.08399555,0.27985346){\makebox(0,0)[lb]{\smash{0.4}}}%
    \put(0.08399555,0.31692703){\makebox(0,0)[lb]{\smash{0.5}}}%
    \put(0.08399555,0.35394188){\makebox(0,0)[lb]{\smash{0.6}}}%
    \put(0.08399555,0.39101545){\makebox(0,0)[lb]{\smash{0.7}}}%
    \put(0.08399555,0.42803029){\makebox(0,0)[lb]{\smash{0.8}}}%
    \put(0.08399555,0.46504514){\makebox(0,0)[lb]{\smash{0.9}}}%
    \put(0.10358152,0.50211871){\makebox(0,0)[lb]{\smash{1}}}%
    \put(0.13394113,0.09082523){\makebox(0,0)[lb]{\smash{0}}}%
    \put(0.25081879,0.09082523){\makebox(0,0)[lb]{\smash{20}}}%
    \put(0.37417552,0.09082523){\makebox(0,0)[lb]{\smash{40}}}%
    \put(0.49755182,0.09082523){\makebox(0,0)[lb]{\smash{60}}}%
    \put(0.62090855,0.09082523){\makebox(0,0)[lb]{\smash{80}}}%
    \put(0.73778621,0.09082523){\makebox(0,0)[lb]{\smash{100}}}%
    \put(0.05773478,0.21705936){\rotatebox{90}{\makebox(0,0)[lb]{\smash{Cumulative Fraction}}}}%
    \put(0.34763288,0.04022587){\makebox(0,0)[lb]{\smash{Average Throughput (KB/s)}}}%
    \put(0.89060214,0.33186216){\makebox(0,0)[lb]{\smash{Geo}}}%
    \put(0.89060214,0.28545168){\makebox(0,0)[lb]{\smash{Random}}}%
    \put(0.89060214,0.23904121){\makebox(0,0)[lb]{\smash{Default}}}%
    \put(0.89060214,0.19257202){\makebox(0,0)[lb]{\smash{Comp.}}}%
  \end{picture}%
\end{tiny}
\endgroup%